\documentclass[a4paper,10pt]{article}
\usepackage{amsmath,amssymb}
\usepackage[dvips]{graphicx}
\title{ SPACE ROTATION THEORY } 
\author{Andrey V. Novikov-Borodin \\
{\it Affiliation:} INR of Russ.AS~~117312 Moscow, 60-th Oct.Ann.pr. 7a, Russia. \\{\it web:}~http://www.inr.ru/$\sim$novikov ,~{\it e-mail:}~ novikov@inr.ru.}
\date{\today}
\begin{document}
\maketitle
%
%
\begin{abstract}

 The theory of Space rotations is introduced. The relativity principle is generalized to satisfy to reference frames rotating in 3D space. It is shown that the most postulates and limitations of quantum theories are consequences of this extended relativity principle. The mathematical, physical and philosophical aspects of the Space rotation theory are discussed. \\
\\ {\bf PACS:} {\it  03.65.Bz -- quantum mechanics, foundations;\\ 03.30.+p -- special relativity; 11.30.Cp -- Lorentz and Poincare invariance;
\\ 12.60.-i -- models of particles and fields beyond the standard model;
\\ 04.20.Gz -- spacetime topology, causal structure, spinor structure;
\\ 04.50.+h - alternative theories of gravity. }

\end{abstract}
\frenchspacing
\tableofcontents
\section*{\bf Introduction}
Can relativistic and quantum theories, these cornerstones of modern physics, actually fit together? May quantum physics become, following by M.Gell-Mann \cite{GM81}, the real theory, but not ``an anti-intuitive discipline ... full of mysteries and paradoxes, that we do not completely understand, but are able to use ... limits, in which, as we suppose, any correct theory should be included''? \\
In spite of all successes of both relativistic and quantum theories, there are big ``clouds on the horizon'' \cite{Butt99} in their correlation. A lot of experiments have confirmed these theories separately, but it looks like their backgrounds contradict to each other. For example, such effects as ``'wave function collapse'', EPR-like effects contradict to the second postulate of Einstein's relativistic theory \cite{AE05}. 
J.Butterfield \cite{Butt99} ``... admits that the physics community at large does not worry so much about this cluster''. But the physics community had worried about it so much before, from the beginning. A.Einstein could not believe in the validity of quantum mechanics \cite{AE35} because of the EPR-like effects. He has seen its incompatibility to the basic relativistic principles, which were confirmed experimentally. But further experiments \cite{ Bell64, Asp82, Chi93} also have confirmed the quantum mechanics conclusions. So, different experiments contradict to different theories, or more correctly to say that the physical backgrounds of these theories contradict to each other! \\
There is more fundamental contradiction. From the special relativity point of view, the description of the quantum mechanical objects contradicts to the causal principle -- the basic one in natural science, the background of our worldview. This is a very serious problem. It is not only physical, but philosophical, worldview question and that is why the physical community prefers to keep silence in this question, to keep all ``as is''. The usual, ``official'' way to solve the problem in modern physics is do not solve it at all by declaring duality of particles, postulating uncertainty relations etc. The excuse is that on this level of investigation, the nature so differs from our everyday views that we cannot even imagine it. Do we know more than the ancient people explaining all events in Nature by wills of their Gods?! I am sure, we need to search the answer, this is a way of Science.  The adepts of ``official'' views transform very easily ``can not imaging'' to ``may not''. And this is quite dangerous, because science is transformed to religion. Unfortunately, adepts of such views may be met at any level of ``the scientific hierarchy'' in physics, even at very high. Such ``religious'' scientific communities may be really dangerous for investigators of this field of science. \\
Realizing all problems and difficulties mentioned above, we are going to introduce {\it the Space rotation theory} (SRT). It seems to us, that this theory can overcome fundamental contradictions between the special relativity and quantum mechanics.  As far as the worldview questions are involved, we will introduce the system of {\it the Physical reality description} (PhRD) from the SRT point of view. After that, we will describe all involved to the PhRD components: the SR frames of references; the physical objects (SR objects), which the SRT operates with; the physical laws and the space-time properties. We will analyze the correspondence between the Space rotation theory with the special relativity, quantum mechanics and, briefly, with some modern theories.  
 
\section{\bf Physical Reality Description}
\label{sec:PhRD}

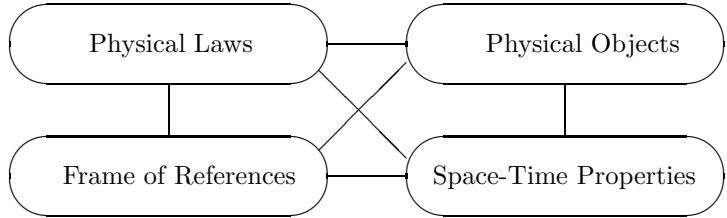
\begin{figure}[t]
	\begin{picture}(450,100)
	\put(100,30){\oval(120,30)}
	\put(60,27){Frame of References}
	\put(100,80){\oval(120,30)}
	\put(70,77){Physical Laws}
	\put(250,30){\oval(120,30)}
	\put(200,27){Space-Time Properties}
	\put(250,80){\oval(120,30)}
	\put(220,77){Physical Objects}
	\put(100,45){\line(0,0){20}}
	\put(160,30){\line(1,0){30}}
	\put(250,45){\line(0,0){20}}
	\put(160,80){\line(1,0){30}}
	\put(157,40){\line(1,1){33}}
	\put(157,70){\line(1,-1){33}}
	\end{picture}
	\caption{The System of the Physical Reality Description. }
	\label{fig:PhRD}
\end{figure}

Notwithstanding the incompatibility of the theory of special relativity and quantum mechanics, the possibility to coincide these theories, apparently, exists. It is based on the generalization, adaptation of the idea of Relativity \cite{AE05} onto reference frames rotating in 3D-space. Such adaptation bases on the idea of the equivalence of the SR frames. The equivalence is understood from point of view of constructing the system of the Physical reality description. Indeed, it concerns with not only the physical and mathematical questions, but also the philosophical, worldview ones. From the first view, it is well-known that physical laws are not invariant related to space rotations \cite{Ryd85}. The problem, however, may be brought to question: what really needs to be included into the system of the physical reality description? It seems strange, but the revolution idea, which lets overcome contradictions mentioned above, is the idea to include the physical object by itself into this system. It means that in the physical Space rotation theory the physical object ``perception'' may be changed in different SR frames. So, for example, the plane electromagnetic wave may be represented in the ``slow rotating'' frame as a set of the spherical waves \cite{Ryd01, Mash00}. The rotating frame analysis in these works is limited by the case of ``slow rotating frames'' \footnote{To introduce the Nonlocality in the object description, the general relativistic principle of Locality during coordinate transformations is still used, so, in general, some initial postulates of these works are under question.}. In general, as it will be shown below, such a modification of the object ``perception'' in different SR frames may be more considerable and has a qualitative character \cite{NB03,NB01}. \\
The idea of Relativity introduced in special relativistic theory has changed our imagination about {\it space-time properties}, has declared the equivalence of the inertial {\it reference frames} from the point of an invariance of {\it the physical laws} in it. The new type of {\it physical objects} was introduced in quantum mechanics, that has caused the introducing of new postulates, new mathematical description, in other words, new {\it physical laws}. These examples show us the interconnection between PhRD components, so we can tell about the {\it system} of PhRD. \\
So, the components, we are going to include to the so-called ``system of the Physical reality description'' are well known, but usually they were not used in ``system''. We will consider that PhRD consists of four interconnected components: {\it the physical laws, physical objects, frames of references and space-time properties} (see Fig. \ref{fig:PhRD}). \\
Unfortunately, there are no complete correspondence between quantum mechanical objects and laws with relativistic imagination of space-time properties. These ``clouds on the horizon'' are principal. In introduced terminology we can say that PhRD's of special relativity and quantum mechanics do not correspond to each other. \\
The special relativity idea of negation of the absolute reference frame leads to its logical extension. Indeed, how to recognize the ``initial'' inertial frame (or a system of inertial frames)? In other words: has the ``inertiality'' an absolute character, and if yes, but why? The idea of relativity is ``selfextracting'', so is needed to be continued or limited. Such efforts to extract the idea of relativity to non-inertial frames (but not rotating ones) were made before. For example, A.Logunov \cite{Log89} generalized the Poincar\'e's relativity principle \cite{Poin04} in the form of: ``For any physical frame of references (inertial or non-inertial), it is always possible to select the system of other frames, in which all physical processes are going similar to the initial frame, so we do not have and it is impossible to have any experimental possibilities to distinguish \ldots what, exactly, frame from this infinite system we are in''. \\
The model of the local, material point (or the point mass) used for the physical object description in classical theories meets also the conceptual problems \cite{Mash00}. We will also point to some additional problem on this way. We will call it the {\it scale problem}. Usually, the continuous space-time is considered in the relativistic theories. From the mathematical point of view, on this way, any local region in space-time has the same {\it power of set} (by Cantor) as the whole space-time. It is not clear, why do we have some scale of the observed physical world. Some additional postulates need to be introduced to fix the problem. One of the possibilities is the time or space discreteness. Partially, the SRT gives the answer to this question by explaining the so called visual discreteness. It will be described below in Sect.\ref{sec:Law}. \\
From the introduced PhRD point of view, the PhRD of the quantum physics is not complete, and not compatible with the relativistic one. There are big problems in the relativistic PhRD too. 

\section{\bf Space Rotation Frames}
\label{sec:Frame}

It was mentioned above in Sect.\ref{sec:PhRD} that the idea of relativity is ``self extracting'', so should be continued or limited. As far as we want to extend it to Space rotations, we need to declare the {\it Equivalence} of the SR frames. So, we are introducing the basic principle of the Space rotation theory -- the principle of the Equivalence of the rotating frames of references. The Equivalence is considered from the point of view of creation of PhRD. The ``mathematical'' rotating frames are equivalent to each other, but what does it mean from the physical point of view? We will try to investigate this question here. 

\subsection{Types of Space Rotations}
We will consider three types of ``mathematical'' 3D-space rotations: the axis space rotation (ASR), the multiple space rotation (MSR) and the sum space rotation (SSR).\\
We will call the space rotation (SR) of the $K'$ frame of references related to $K$ in time $t$ with the frequency $\omega$ about some fixed space axis as an {\it axis} space rotation (ASR). We will also use notations: $\tau = c t$, $\Omega=\omega/c$, where $c$ is a speed of light. Space coordinates in $K'$ will be marked by apostrophes as $(x',y',z')=X'$. The coordinate transformation between the ``initial'' frame $K$ and rotating one $K'$ will be written as: 
\begin{equation}
X'=X\cdot (A^{ASR}_{z})^{\pm},\qquad (A^{ASR}_{z})^{\pm}=\left (
\begin{array}{ccc} \cos(\omega\,t) & \mp\sin(\omega\,t) & 0 \\
 \noalign{\medskip} \pm\sin(\omega\,t) & \cos(\omega\,t) & 0 \\
 \noalign{\medskip} 0 & 0 & 1
\end{array}\right ).
\label{ASR}
\end{equation}
Here $(A^{ASR}_{z})^{\pm}$ is a transformation matrix of the ASR about $z$-axis in space (without loosing the generality). At any time the functional determinant of the ASR transformation matrix is equal to unit ($\det A = 1$).\\
We will call the combination of ASR rotations with transformation matrices $A_1, A_2, \dots, A_n$ by definition as a {\it multiple} space rotation (MSR) and a {\it sum} space rotation (SSR), if the general transformation matrices of these rotations $A^{MSR}$ and $A^{SSR}$ will be expressed as: 
\begin{equation}
A^{MSR}=\, \prod _{k=1}^{n}A_{{k}},\qquad A^{SSR}= \sum _{k=1}^{n} A_{{k}}. \label{MSSR}
\end{equation}
Let the $K$ frame rotates related to $K'$ with the transformation matrix $A_{1}$ and $K''$ frame rotates related to $K'$ with the transformation matrix $A_{2}$. The transformation between $K''$ and $K$ will be expressed by the transformation matrix $A^{MSR}=\{a_{ij}\}=A_1 \cdot A_2=\sum _{k=1}^{n}a^{1}_{ik}a^{2}_{kj}$. The SRs in SSR are considered from the $K$ frame of references. \\
It may be shown that some MSR and SSR properties in Table \ref{tab:SRProp} are true.
\begin{table}
\begin{center}
\caption{The MSR and SSR properties}
\begin{tabular}{|l|c|c|}
\hline
&MSR&SSR \\
\hline
1&$\left ( A^{MSR}\right )^{-1} = \left (A^{MSR}\right )^{T}.$ & $\left ( A^{SSR}\right )^{-1} \neq \left (A^{SSR}\right )^{T}.$ \\
2&$ \det A^{MSR} =\, \prod _{k=1}^{n} \det  A_{{k}} =1.$ & $\det A^{SSR}\neq 1.$ \\
3&$A^{MSR}_{..ij..}\neq A^{MSR}_{..ji..}.$&$ A^{SSR}_{..ij..}=A^{SSR}_{..ji..}. $\\
\hline
\end{tabular}
\end{center}
\label{tab:SRProp}
\end{table}
The MSR transformations (and the ASR as a particular case) are forming the rotation group like ${\cal O}(3)$, where $t$ looks like parameter. Let's denote this group as an extended rotation group ${\cal \widehat O}(3)$. Mathematically, MSR transforms one sphere in $K$ to another in $K'$. In case of normalized MSR, for any $t$: $\|X'\|^2=\|X\|^2$. The SSR transformations do not form a group. In distinguish to MSR, the SSR transforms the sphere in $K$ inside the full sphere in $K'$. The value $\|X'\|^2$ is not constant and depends on $t$. \\
Due to matrix properties, one can conclude that the transformation matrix of any space rotation with at least one common rotation point may be represented as a sum of products of the ASR transformation matrices: 
\begin{equation}
X' = X \cdot A^{SR}= X \cdot \left \{ \sum_{i} A^{MSR}_{i} \right \}. \label{SR}
\end{equation}
The expression for the coordinate transformation without common rotation point one can get by replacing space coordinates $X$ by $X+X_0$ on the corresponding step, where $X_0$ is the origin coordinates displacement on this step. 

\subsection{Space Rotation Equivalence and Metrics}
The first of all, the SR frames Equivalence means that Minkowski space-time can be built in any SR frame, so ``internal'' metrics can be introduced inside any SR frame. For some space rotation with the transformation matrix $A$ between $K$ and $K'$ we will analyze the expression for ``interval'' in rotating frame $K'$ in the form: 
\begin{equation}
ds'^{2}=\, c^{2}dt^{2}-\, \|dX'\|^{2}, \label{Interval}
\end{equation}
where $ dX'$ is defined as $dX' = dX A+X dA$ and $\| dX'\|^{2}=dX' dX'^{T}$ \footnote{Here $t'=t$ is assumed, because it is so in both SR frames at least at one point  - the rotation point. We may consider it as an interval seen in $K'$ from the rotation point. It may be considered as a first order approximation to the real interval, but we believe that main conclusions will be the same.}. 
So, for (\ref{Interval}) we will get: 
\begin{eqnarray}
\left (ds'_{SR}\right )^{2} & = & c^{2}dt^{2}- X dA dA^{T} X^{T}- dX A A^{T} dX^{T}- \label{X2}\\ 
& & -\left ( dX A dA^{T} X^{T}+ X dA A^{T} dX^{T} \right ) dt. \nonumber
\end{eqnarray} 
For the ASR and MSR this expression can be simplified:
\begin{eqnarray}
\left (ds'_{MSR}\right )^{2} & = &  \left [ c^{2}- \left ( X \frac {\partial A}{\partial t} \frac {\partial A^{T}}{\partial t} X^{T}\right )\right ] dt^{2}-  \|dX\|^{2}- {}\label{X} \\
 & & {} - \left \{ dX\, A \frac {\partial A^T}{\partial t} X^{T}+ X \frac {\partial A}{\partial t} A^{T} dX^{T} \right \}. \nonumber
\end{eqnarray}

$\bullet$ {\it  Metrics in rotation point.}\\ 
The rotation point by definition is a common point of the axes of rotation, included to the considered space rotation. If we consider the interval (\ref{Interval}) from the rotation point $X'_{rp}=\, X_{rp}=(0,0,0)$ (the same in $K$ and $K'$), we will get for any time $t$ (the transformation matrix is normalized): 
\begin{equation}
\|\Delta X'\|^{2}= X'\cdot X'^{T}= X\cdot A^{MSR}(t)[A^{MSR}(t)]^{T}\cdot X^{T}= 
X\cdot X^{T}= \|\Delta X\|^{2}. \nonumber
\end{equation}
It means, according to (\ref{Interval}), that $K$ and $K'$ are invariant in rotation point and this invariance has a local (or even point) character. The situation with points differed from the rotation one is another.\\

\begin{figure}[t,s]
	\begin{center}
		\includegraphics*[angle=270,width=50mm]{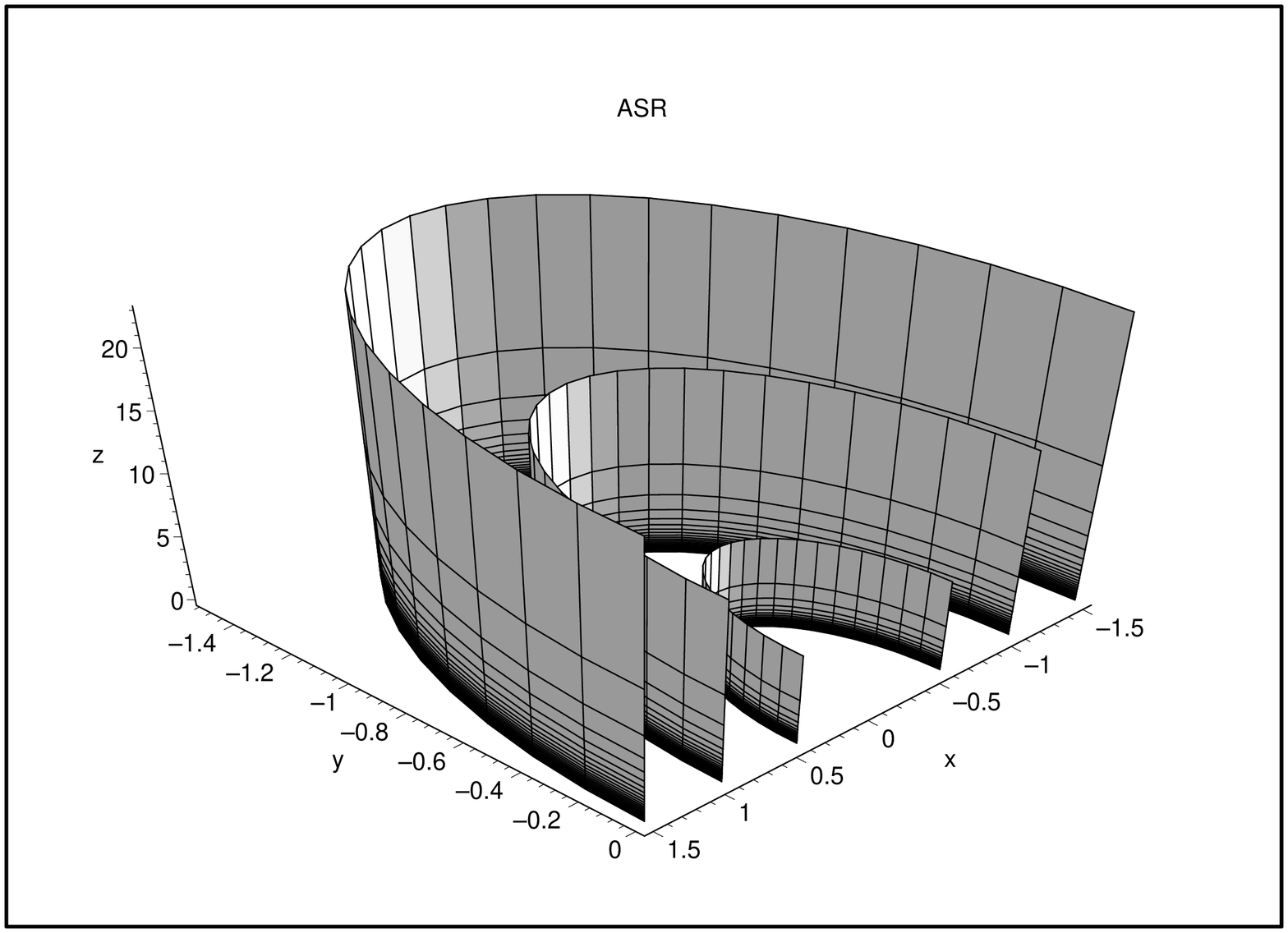}
		\includegraphics*[angle=270,width=50mm]{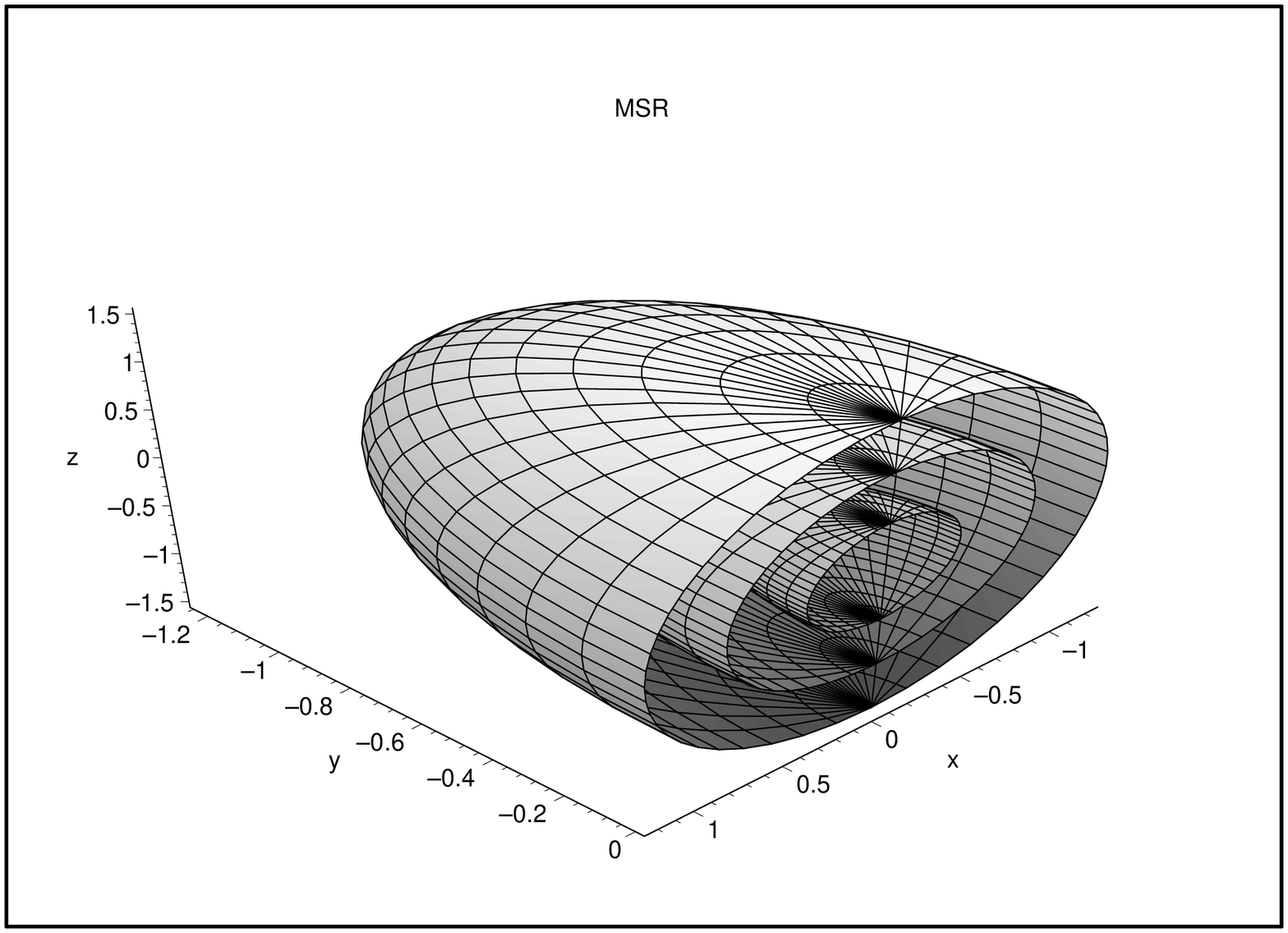}
	\caption{The ASR and MSR stable regions.}
	\label{fig:SRobj}
	\end{center}
\end{figure}

$\bullet$ {\it  ASR metrics} \\ 
Let's consider the transformation matrix $(A^{ASR}_{z})^{\pm}$ from (\ref{ASR}) without loosing the generality. From (\ref{X2}), we will get the expression: 
\begin{eqnarray}
\left (ds'_{ASR}\right )^{2}&=&\left [ c^2- \omega^{2} \left (x^{2}+y^{2}\right )\right ] dt^{2}\mp 2\,\left (y dx-x dy\right )\omega dt- \label{ASRintx}\\
& & - ( dx^{2}+dy^{2}+dz^{2} ), \nonumber
\end{eqnarray} 
that in cylindrical coordinates $(\rho,\phi,z)$ will be: 
\begin{equation}
\left (ds'_{ASR}\right )^{2}=\, \left (c^2- \omega^{2}\rho^{2}\right ) dt^{2}\pm\, 2\rho^{2}\omega\, d\phi\, dt-\, \rho^{2}\, d\phi^{2}- d\rho^{2}- dz^{2}. \label{ASRint}
\end{equation}
In spite of our assumption ($t=t'$) in getting (\ref{Interval}) is under question, this expression is completely the same as it is usually used in handbooks for the description of metrics in such rotating frames (see, for example, Landau \cite{Land85}). It is a good sign on our way. \\
Physically, it is clear that on some distance from the rotation axis ($\rho^2>c^2/\omega^2$), the SR object in $K'$ will have the speed in $K$ more than the speed of light. It is also impossible for the real physical object in our PhRD. So, if SR object exists in $K'$, it needs to be confined inside some surface. This surface, defined by the equation $\rho^{2}=\, c^2/ \omega^{2} \leftrightarrow g_{00}=0$ ($g_{00}$ is the metric tensor coefficient at $dt^2$), divides the space into the {\it internal} and {\it external} regions correspondingly. This surface looks like the horizon for real physical objects in $K'$ from the $K$ point of view. We will call this surface as a horizon surface \footnote{In fact, the physical object will be confined inside the horizon of events $ds^2=0$. In our case, for simplicity, $g_{00}=0$ $ds^2 < 0$, so the horizon of events are inside the horizon surface.}. The equation $g_{00}=0$ with different frequencies $\omega$ defines a set of cylindroids about the $z$-axis (see Fig.\ref{fig:SRobj}). \\
The metric tensor corresponding to the ASR interval has non-diagonal elements and, so, SR frames are immeasurable. It means that we could not write an analytical expression describing SR objects in rotating frames. \\

$\bullet$ {\it  MSR metrics} \\
Let's consider the MSR $ A^{MSR}=\, A^{ASR}_{z}(\omega_{1})\, A^{ASR}_{x}(\omega_{2})\, A^{ASR}_{y}(\omega_{3})$, where $\omega_{1},\, \omega_{2}$ and $\omega_{3}$ are rotation frequencies. For the MSR the analysis is differed from the ASR one, because the interval (\ref{X}) depends on time and is complicated. For example, for $A^{ex}=\, A^{ASR}_{z}(\omega)\, A^{ASR}_{x}(\omega)$, even $\|dX'\|^{2}$ will be expressed as: 
\begin{eqnarray}
\|dX'\|^{2} &=& \big [2\,{x}^{2}-2\,x\sin(\omega\,t)y\cos(\omega\,t)+{y}^{2}-{x}^{2}
\cos^{2}(\omega\,t)-{} \nonumber\\
 & & {}-2\,yz\sin(\omega\,t)+{y}^{2}\cos^{2}(\omega\,t)-2\,xz\cos(\omega\,t)+{z}^{2}\big ]{\omega}^{2}{{\it dt}}^{2}+ {} \nonumber\\ 
 & & {}+2\, \big [\left (y-\,z\sin(\omega\,t) \right ){\it dx}+\left (z\cos(\omega\,t)-\,x\right ){\it dy}- {} \nonumber\\
 & & {} -\left (\cos(\omega\,t)y-\,\sin(\omega\,t)x\right ){\it dz}\big ]\omega\,{\it dt}+{{\it dx}}^{2}+{{\it dy}}^{2}+{{\it dz}}^{2} \nonumber.  
\end{eqnarray}
We will analyze the interval, averaged in time, from the point of view of the observer in $K$ and will consider that the period of time of the observation is much longer than the period of any rotation included in the MSR. We consider that such situation is quite close to the real situation in particle physics. We will use the following expression for ``averaging'': 
\begin{equation}
\langle f \rangle_{t}=\, \lim _{t\to \infty}\frac {1}{2t}\int_{-t}^{+t}\, f(t)\, dt. \label{Average}
\end{equation} 
On this way, for the MSR interval one can get: 
\begin{eqnarray}
 & & \langle \left (ds'_{MSR}\right )^{2} \rangle_{t}= \bigg \{ c^2-\, \bigg [\left (x^{2}+y^{2}\right )\omega_{1}^{2}+\left (\frac {1}{2}\,x^{2}+\, \frac {1}{2}\,y^{2}+z^{2}\right )\omega_{2}^{2}+ \label{MSRint}\\
 & & +\left (\frac {3}{4}\,{x}^{2}+\frac {3}{4}\,{y}^{2}+\frac {1}{2}\,{z}^{2}\right )\omega_{3}^{2}\bigg ]\bigg \}dt^{2}- 2\,\left (y\,dx-\,x\,dy\right )\omega_{1}\,dt - dx^{2}- dy^{2}- dz^{2}. \nonumber
\end{eqnarray} 
On the analogy of the ASR analysis, we can conclude that there are some regions localized in space, defined by the equation $g_{00}=0$, where the physical object will be stable in time from the $K$ point of view. In spherical coordinates $(r,\theta,\phi)$, these stable regions will satisfy the equation: 
\begin{eqnarray}
r^{2} = \bigg [\left (\frac {\omega_{1}}{c}\right )^{2}\, \sin^{2}(\theta)+\, 
\left (\frac {\omega_{2}}{c}\right )^{2}\,\left (1-\,\frac {1}{2}  \sin^{2}(\theta)\right ) 
+\frac {1}{2}\left (\frac {\omega_{3}}{c}\right )^{2}\,\left (1+\,\frac {1}{2}  \sin^{2}(\theta)\right )\bigg ]^{-1}.
\label{MSRobj}
\end{eqnarray} 
This equation for different frequencies describes ellipsoids in space (see Fig.\ref{fig:SRobj}). \\

$\bullet$ {\it  SSR metrics}\\
 For the SSR with $A^{SSR}=\, A^{ASR}_{z}(\omega_{1})+\, A^{ASR}_{x}(\omega_{2})+\, A^{ASR}_{y}(\omega_{3})$, we need to use the expression (\ref{X2}) with ``averaging'' (\ref{Average}). This way, one can get: 
\begin{eqnarray}
 & & \langle \left ( ds'_{SSR}\right )^{2} \rangle_{t} = \left \{ c^2-\, \left [\left (x^{2}+y^{2}\right )\omega_{1}^{2}+\left (y^{2}+z^{2}\right )\omega_{2}^{2}+\left (x^{2}+z^{2}\right )\omega_{3}^{2}\right ]\right \} dt^{2}- {} \label{SSRint}\\
 & & {} -2\, \left [\left (y\,dx-x\,dy\right )\omega_{1}-\left (y\,dz-\,z\,dy\right )\omega_{2}+\left (z\,dx-\,x\,dz\right )\omega_{3}\right ] dt - \nonumber
 - 3 \left ( dx^{2}+\,dy^{2}+\,dz^{2}\right ). \nonumber
\end{eqnarray} 
It is also possible to separate the space onto the internal and external parts and, also, like in the MSR, the horizon of events surface has an ellipsoidal structure (see Fig.\ref{fig:SRobj}), but the ``internal'' space characteristics will differ from the MSR ones. We will discuss it later in Sect.\ref{sec:Object}. 

\section{\bf Space Rotation Objects}
\label{sec:Object}

It was mentioned in Sect.\ref{sec:PhRD} that ``visual'' characteristics of the SR objects may differ considerably from the ``initial'' characteristics of these objects in non-rotating frame. Here we will try to find the most general properties of the SR object, which are defined by the metrics of the SR frames. \\ 
The first of all, the ``visual'' metrics structure (Fig.\ref{fig:SRobj}) leads to the conclusion that from the observer point of view any physical object in SR frame can not escape from the internal region, because it is limited by the horizon surface. This way it means that the SR object is localized in space and it is possible to tell about ``sizes'' of SR objects. For MSR and SSR, these objects are localized in 3D-space, and for ASR -- in 2D-space. The conclusion about the existence of the SR stable regions is an additional characteristic to the ``mathematical'' model, the ``physical'' aspect of the extended rotation group ${\cal \widehat O}(3)$. Anyway, the existence of the stable objects is not a problem (as in modern physics), but a direct consequence of the SRT. \\ 
The horizon surface is a very specific region. This way, a lot of characteristics of the SR objects are hidden for the external observer. The question: Can we get complete information about the SR objects? -- Is open. \\

$\bullet$ {\it  ASR object.}\\
For the ASR, the physical object in $K'$ from $K$ will be observed as an object, localized in the internal region, stable on its edge and freely movable along with the rotation axis. As far as the rotation frequency with the same value may have two direction of rotation (may have positive or negative sign), it may be two different types of these objects. One can say that this object will have an additional characteristic from the $K$ point of view. This characteristic looks like {\it spin} that, on this approach, may be directed along or opposite the rotation axis. In elementary particle physics, this behavior of the ASR physical object has some correspondence with {\it neutrino}. Also,
analyzing Eq.(\ref{ASRint}) one can find that it needs to be for the ASR object the correlation between transverse and longitudinal motion: 
$dz^2/dt^2=c^2-\omega^2\rho^2-d\rho^2/dt^2$. Here $d\phi=0$, so as we consider the rotation as a hidden parameter. It may be considered as neutrino oscillations. \\

$\bullet$ {\it  MSR object.}\\
The MSR object is localized in 3D-space. As far as the ``initial'' SR object has to have some energy in $K'$ (it may be non-localized in $K'$), the corresponding object in $K$ with some energy is localized for the observer. For him, we may suppose, this object needs to have characteristics like rest mass of a particle, because of well-known Einstein relation $E = m_0 c^2$ (see Sect.\ref{sec:Law} in detail). \\ 
Comparing expression (\ref{MSRint}) with (\ref{ASRintx}), (\ref{ASRint}), one can see that although the MSR consists of a few orthogonal ASRs, the MSR object, localized in space, anyway has some axis, picked out in space. Note, that it is some pseudo-axis, because it is some ``averaged in time'' axis. We can suppose, as before, during the ASR object analysis, that for the observer the MSR object also needs to have the physical ``spin''-characteristic. Here, it is not seen any limitations to the spin direction related to the direction of the particle movement (in distinguish to the ASR object). That also corresponds to the experimental data for the massive particles with spin. Such MSR objects are similar to {\it fermions}. Interesting theoretical and experimental confirmation of this idea may be seen in works of Y.Liu \cite{Liu00} in his model of description of CP violation. \\

$\bullet$ {\it  SSR object.}\\
The SSR object is localized in space and, also, like the MSR object, has an ellipsoidal structure (see Fig.\ref{fig:SRobj}), but it is so only in ``averaging'' in time. It is also possible to separate the space onto the internal and external parts, but again, the ``internal'' characteristics will differ from the MSR ones. The SSR objects are localized in space and, apparently, are massive. But, generally, it is impossible to pick out any space axis (even averaged) for the SSR object ``as a whole''. SSR objects may be ``complex'', i.e. consisted of a few ASR or MSR objects, but their characteristics ``as a whole'' differ from the characteristics of the included SR objects. For the SSR objects $\|X'\|^2\neq\|X\|^2$ (with ``averaging'' some additional coefficients will appear). It looks like due to the SSR metrics changing the intensity of the ``interaction'' between the included ASRs is increasing, that, may be, will correspond to electroweak and strong forces. There are arising the analogies with objects of chromodynamics (QCD). On this approach, if we will continue the analogies between SR objects and particles, it is possible to find some correspondence of the SSR objects with models of the particles consisted of {\it quarks} (the SSR of ASR objects) and even with the {\it nuclei} (the SSR of MSR objects). This way, for example, strange behavior of strong interaction, the confinement looks as logical consequence of SRT. \\
It is only qualitative correspondence between SR and quantum physics objects. But, as it was mentioned before, due to SR frame properties, a lot of SR object characteristics are hidden, because of SR frames are immeasurable and SR objects are inside the horizon surface. On this step, we do not even know how SR objects may be observed from other SR frame. Now, we will try to investigate these questions, to get more possible SR object characteristics to compare it with the quantum physics models. 

\subsection{Quantification}
If the SR object in $K'$ exists, it needs to be a source of some ``influence'' in $K$, otherwise we would not know about it. We consider, that this ``influence'' is well known in $K$ (we don't mean quantum physics here). Furthermore, this influence cannot be an energy source in $K$ or it has to be localized in space, differently, the SR object would be a permanent power source or it would be unstable. This {\it localization conditions} are connected with the energy conservation law in $K$. \\ 
We will analyze the stable and localized SR object. Let's consider that it is a source of some influence $u(X,t)$ in $K$. The source function $\rho (X,t)$, corresponding to SR object, describing some its property, is defined in some internal region ${\cal G}$ in $K$ and is equal to zero in external region. The wave-like influence $u(X,t)$ needs to satisfy the wave equation with the following fundamental solutions \cite{Vlad81} for one-, two- and three- dimensional space: 
\begin{eqnarray}
\nabla^{2} u(X,t)- \frac{1}{v^{2}}\frac{\partial^{2}}{\partial t^{2}} u(X,t)=-\rho(X,t)\qquad \qquad \qquad \qquad \qquad \label{waveeq} \\
{\cal E}_1(X,t)=\frac{\theta (vt-|X|)}{2v},\,{\cal E}_2(X,t)=\frac{\theta (vt-|X|)}{2\pi v \sqrt{v^2t^2-|X|^2}},\,
{\cal E}_3(X,t) = \frac{\theta (t)}{2\pi v}\delta (v^2t^2-|X|^2). \nonumber 
\end{eqnarray} 
Here $v$ is a speed of the influence wave, $\delta$ is the Dirac delta function, $\theta$ is the Heaviside staircase function. Considering the SR object as a source of the wave-like influence with the frequency, corresponding to the rotation frequency $\omega$, and supposing $k=\omega/v$ (with $v=c$, $k= \Omega$), $\rho (X,t)= {\cal P}(X) \exp(\pm \imath\, \omega t)$ (\footnote{The question of the source function expression, we have used, is quite serious, it needs the $omega$--invariance hypothesis, which is introduced and considered in the next Sect. \ref{sec:Law}.}), $u(X,t)= U(X) \exp(\pm \imath\, \omega t)$ ($\imath$ is an imaginary unit), for $U(X)$ we will have the Helmholtz equation with corresponding fundamental solutions \cite{Vlad81}: 
\begin{eqnarray}
\nabla^{2}U(X)+\, k^{2}U(X)=\, -{\cal P}(X), \qquad \qquad \qquad \qquad \qquad \label{DE} \\
{\cal E}_1(X)= \pm \frac{1}{2\imath\, k}e^{\pm\imath\, k |X|},\,{\cal E}_2(X)= \mp \frac{\imath}{4}H_0^{(1),(2)}(k |X|), \,
{\cal E}_3(X) = - \frac{1}{4\pi |X|}e^{\pm \imath\, k |X|}, \nonumber 
\end{eqnarray}
where $H_0^{(1),(2)}$ are the Hankel functions of the first and second kind. Note, that this equation gives the steady-state solutions. \\

\begin{figure}[t]
	\begin{center}
		\includegraphics*[angle=270,width=50mm]{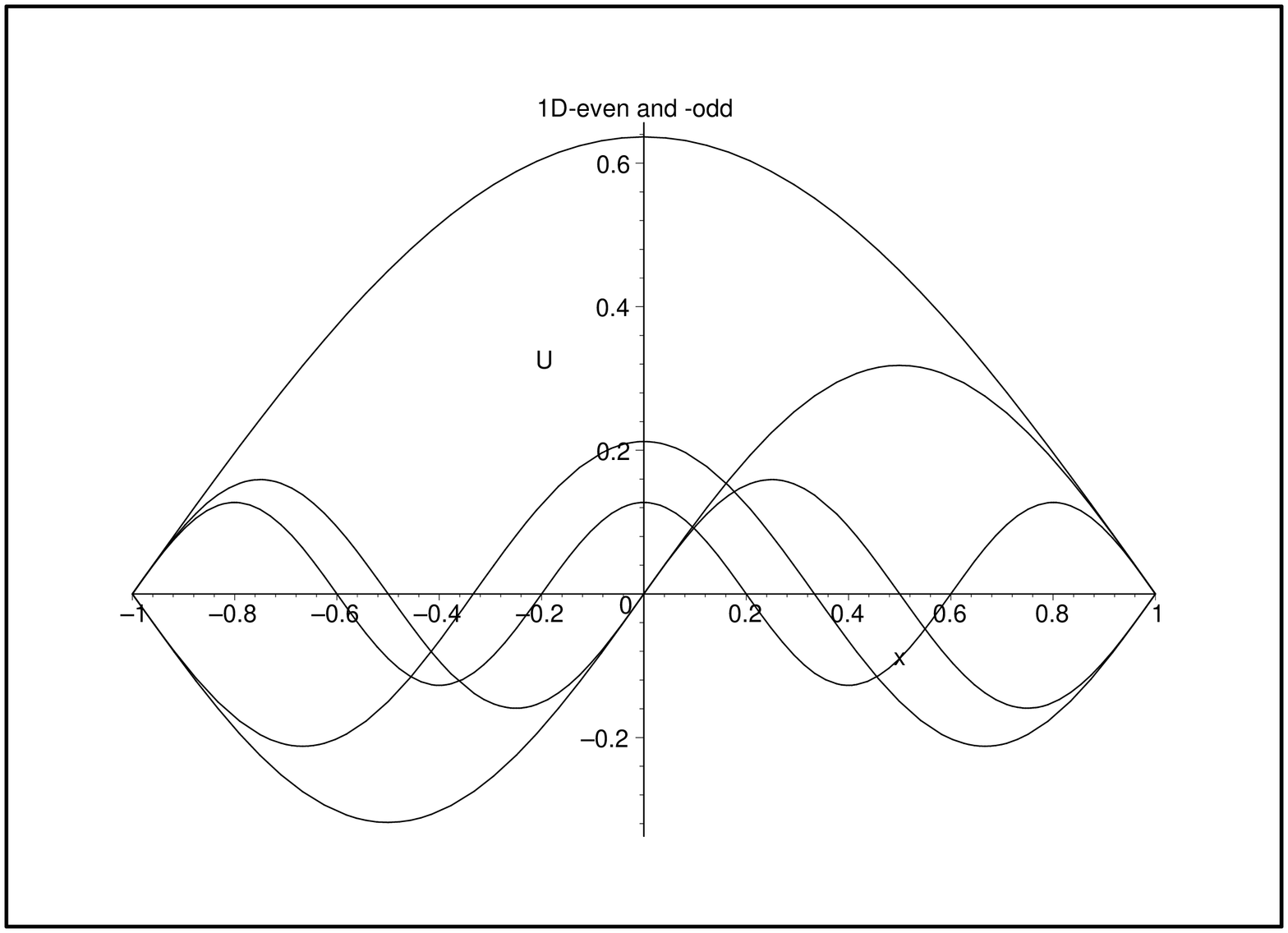}
		\includegraphics*[angle=270,width=50mm]{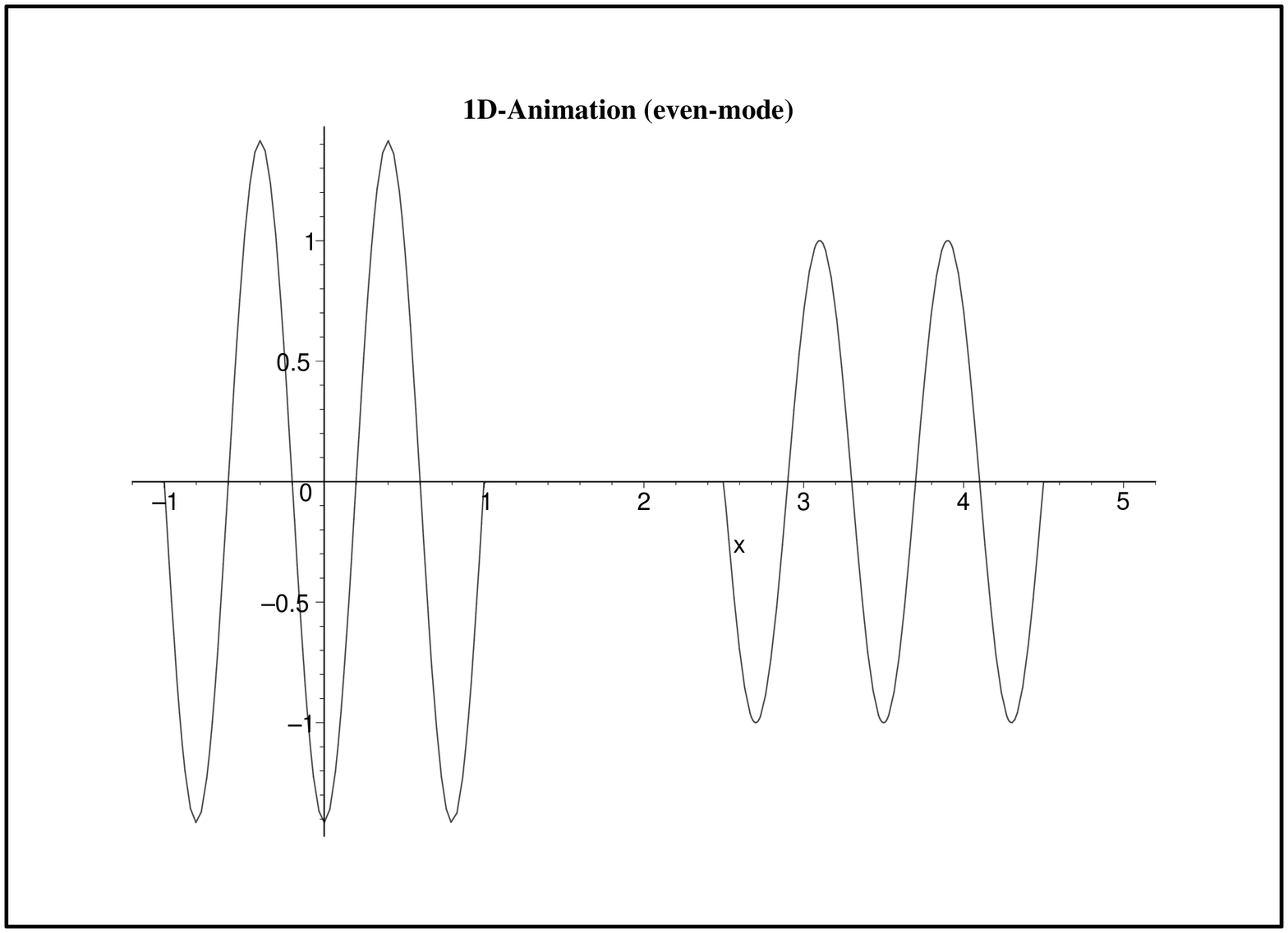}
	\caption{The 1D SR object distribution and establishing.}
	\label{fig:Obj1D}
	\end{center}
\end{figure}

$\bullet$ {\it 1D solutions} \\
In one-dimensional case, the solutions of the Eq.(\ref{DE}) may be represented as: 
\begin{eqnarray}
U(x)&=& C_1 e^{-\imath k x} + C_2 e^{\imath k x} + \label{1D} \\
&+&\frac{1}{2 \imath k} \left[ e^{-\imath k x}\int_{\cal G} e^{\imath k x} {\cal P}(x)dx - e^{\imath k x}\int_{\cal G} e^{-\imath k x} {\cal P}(x)dx \right]. \nonumber
\end{eqnarray} 
From here with ${\cal P}(x)=q[\delta(x-a)\pm\delta(x+a)]$ ($q$ is the source intensity), one can get: 
\begin{eqnarray}
U(x)=\frac{q}{2 \imath k} \left[ e^{-\imath k x}\left( e^{\imath k a} \pm e^{-\imath k a} \right) - e^{\imath k x}\left( e^{-\imath k a} \pm e^{\imath k a} \right) \right]. \nonumber
\end{eqnarray} 
The points $x=a$ and $x=-a$ are the sources of the influence waves. The wave $e^{-\imath k x}$ corresponds to the wave moving along with the $x$-axis and $e^{\imath k x}$ -- to the opposite direction. The localization condition of $U(x)$ may be satisfied if these waves from these points compensate each other in external regions. In the ``internal'' $(-a<x<a)$ and the ``external'' regions $(x< - a)$ and $(x>a)$ these solutions can be represented by even and odd modes: 
\[U^{in}_{even}(x)=\pm\frac{q}{\imath k}e^{\pm \imath ka}\cos(kx),\; U^{in}_{odd}(x)=\pm\frac{q}{k}e^{\pm\imath ka}\sin(kx)\] 
\[U^{ex}_{even}(x)=\pm\frac{q}{\imath k}e^{\mp \imath kx}\cos(ka),\; U^{ex}_{odd}(x)=\pm\frac{q}{k}e^{\mp \imath kx}\sin(ka).\]
It follows from here, that with $ka = -\pi/2 + \pi n$ for even and with $ka = \pi n$ for odd modes, where $n$ is natural, the SR object influence in external regions will be equal to zero at any time, while it won't be zero in the internal region. \\ 
The analysis of the establishing of the found steady-states of the SR object by Eq.(\ref{waveeq}) shows (see Fig.\ref{fig:Obj1D}) that there exist objects spreading out the ``nascent'' SR object with the speed of influence waves. These additional objects represent the wave trains. The number of trains corresponds to the corresponded train number of the steady-state object. So it is possible to tell about the ``families'' or ``classes'' of the steady-state SR objects and corresponding moving objects. In particle physics, analogies with the neutrino families are very transparent. \\

$\bullet$ {\it 3D solutions} \\ 
In the three-dimensional case, if the source function may be expanded in series as ${\cal P}(X)= \sum_n q_n(r)\sum_{l,m}Y^n_{l,m}(\theta, \phi)$ ($Y^n_{l,m}(\theta, \phi)$ are the spherical harmonics), it is possible to separate variables in Eq.(\ref{DE}) and to find the solutions in the form $U(X)= \sum_n R_n(r)\sum_{l,m}Y^n_{l,m}(\theta, \phi)$. The functions $R_n(r)$ are the solutions of the equation:
\begin{equation}
\frac{1}{r^2}(r^2 (R_n)_r)_r+\left[k^2-\frac{l(l+1)}{r^2} \right]R_n+q_n(r)=0, \nonumber 
\end{equation} 
where $R_r = \partial R/ \partial r$. The solutions of this equation for some $n$ may be represented as: 
\begin{eqnarray}
R(r)&=& C_1 h^{(1)}_l(kr) + C_2 h^{(2)}_l(kr) + \label{3D} \\
&+&\frac{k}{2 \imath} \left[ h^{(2)}_l(kr) \int_{\cal G} h^{(1)}_l(kr) q(r) r^2 dr - h^{(1)}_l(kr) \int_{\cal G} h^{(2)}_l(kr) q(r) r^2 dr \right], \nonumber
\end{eqnarray} 
where $h^{(1),(2)}_l$ are the spherical Hankel functions of the first and second kind. \\

\begin{figure}[t]
	\begin{center}
		\includegraphics*[angle=270,width=50mm]{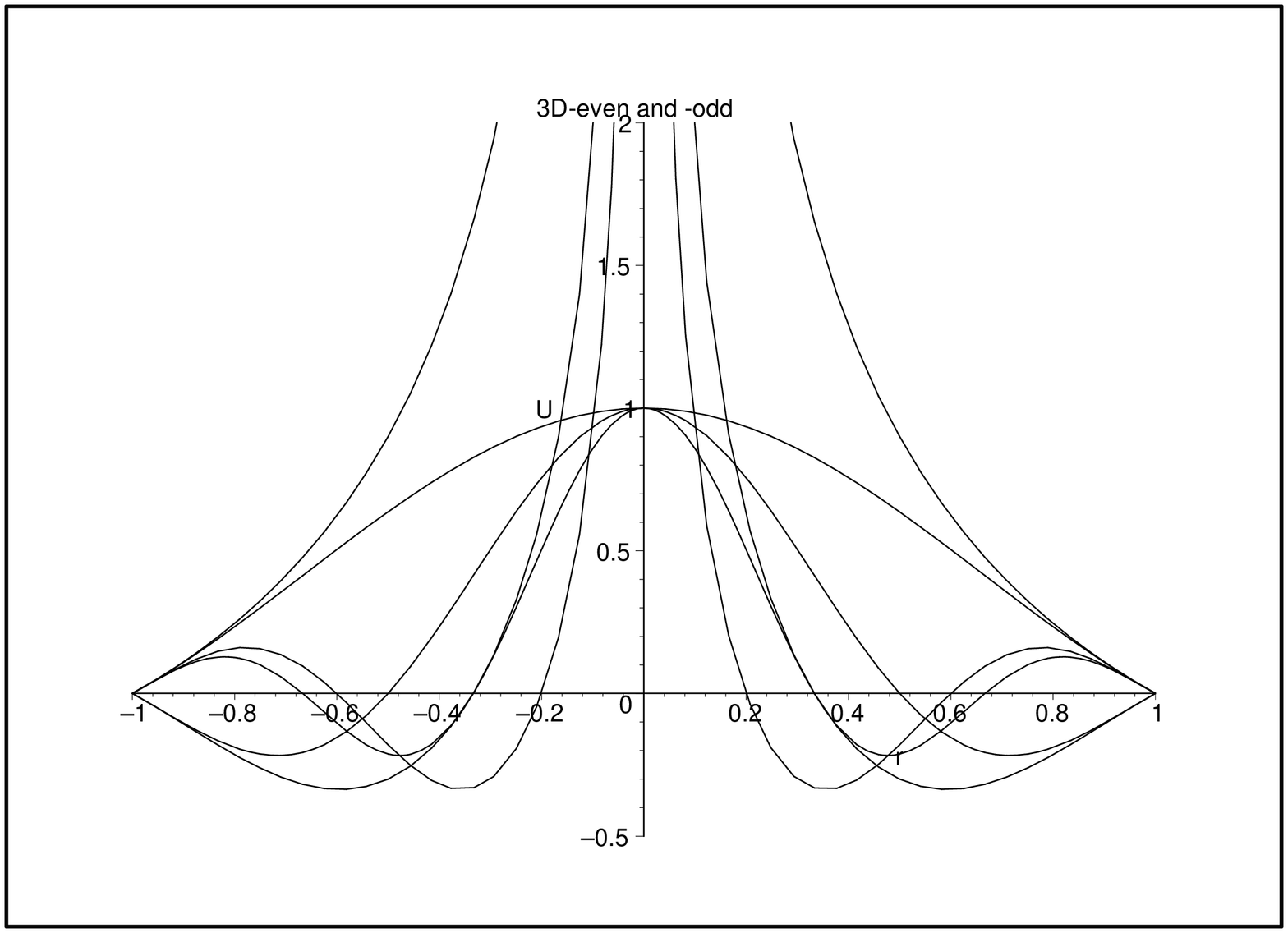}
		\includegraphics*[angle=270,width=50mm]{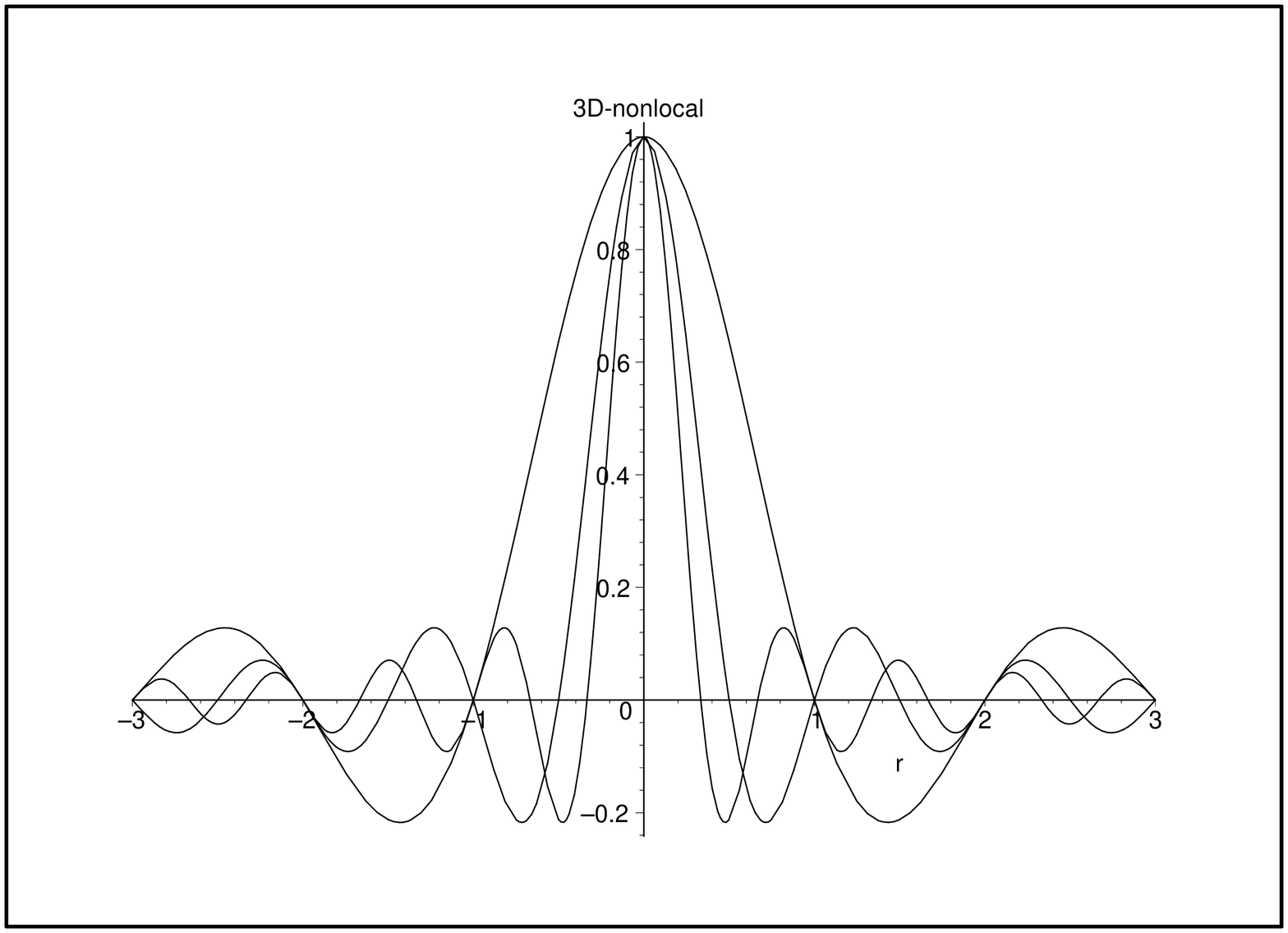}
	\caption{The 3D SR object distribution (localized and nonlocalized).}
	\label{fig:Obj3D}
	\end{center}
\end{figure}

For $q(r)=q\delta(r-a)/(4 \pi a^2)$, where $\left(\int_{\cal G} q(r)r^2 dr = q\right)$, it is also possible to get the steady-state solutions of the Eq.(\ref{DE}). Analogously to 1D case, for internal and external regions, for even- and odd- modes, it may be represented as: 
\[ R^{in}(r)=\frac{qk}{2\imath} \left[ h^{(1)}_l(kr) h^{(2)}_l(ka) \pm h^{(2)}_l(kr) h^{(1)}_l(ka) e^{-2\imath ka} \right], 0<r<a ;\] 
\[ R^{ex}(r)=\frac{qk}{2\imath} h^{(2)}_l(kr) h^{(1)}_l(ka) \left[ 1 \pm e^{-2\imath ka} \right], r>a .\] 
For $l=0$ these solutions can be represented as: 
\[U^{in}_{even}(r)=-\frac{\imath q}{kar}e^{- \imath ka}\cos(kr),\; U^{in}_{odd}(r)=\frac{q}{kar}e^{-\imath ka}\sin(kr) \] 
\[U^{ex}_{even}(r)=-\frac{\imath q}{kar}e^{- \imath kr}\cos(ka),\; U^{ex}_{odd}(r)=\frac{q}{kar}e^{-\imath kr}\sin(ka). \]
It follows from here, that with $ka = -\pi/2 + \pi n$ for even and with $ka = \pi n$ for odd modes, where $n$ is natural, the SR object influence in external regions will be equal to zero at any time, while it won't be zero in the internal region (Fig.\ref{fig:Obj3D}). So, the localization conditions for 3D SR objects may be satisfied. We have found that the SR objects satisfy to some {\it quantification conditions}. Remarkable, that the quantification takes place without any external force field. \\

$\bullet$ {\it Electric charge} \\
Let us consider the particular solution of Eq.(\ref{DE}) in 3D-space for ${\cal P}(r)=q\delta(r-a)/(4 \pi a^2)$. We consider that the influence is an electromagnetic wave. From Expr.(\ref{3D}) with $l=0$ one can get (SI): 
\begin{equation}
U(r)=\frac{q}{4\pi \epsilon_0 a} \frac{sin[k(r-a)]}{kr}, 
\label{Q}
\end{equation} 
where $\epsilon_0$ is a permittivity of vacuum. The corresponding distributions are shown on Fig.\ref{fig:Obj3D}\\
This solution does not satisfy to the localization condition, it is defined in a whole $3D$-space. The only way to overcome the contradiction with the energy conservation law is to demand the wave frequency to be equal to zero. This way there will not be an energy flow from or to SR object. The limit of Eq.(\ref{Q}) with $k$ tends to zero is: 
\begin{equation}
\lim_{k \rightarrow 0}U(r)=\lim_{k \rightarrow 0}\frac{q}{4 \pi \epsilon_0 a}\frac{sin[k(r-a)]}{kr}=\frac{q}{4 \pi \epsilon_0}\left( \frac{1}{a}-\frac{1}{r} \right). 
\label{Q0}
\end{equation} 
Here $U(r)$ looks like a potential of particle with the charge $q$. The additional constant $q/(4 \pi \epsilon_0 a)$ in the potential expression (\ref{Q0}) usually does not play any role in physics, but here it corresponds to the rest mass of a particle. Indeed: 
\[m_0={\cal E}/c^2=qU(\infty)/c^2=q^2/(4 \pi \epsilon_0 c^2 a)=\mu_0 q^2/(4\pi a). \]
This expression is exactly the relation between the electron rest mass $m_0$ and its classical radius $a$. \\
It looks strange, but the first term in Expr.(\ref{Q0}) corresponds to the electron rest mass and the second one corresponds to its electrostatic potential. Is it accidental or one of the basic properties? We will not comment it in this paper. 

\subsection{DeBroglie wave.} 
The SR object influence, taking into account the {\it quantification condition} $k_n=\omega_n/ v$, may be represented as: 
\begin{equation}
u(X,t) = \sum_n U(\omega_n,X) \exp(\imath \omega_n t). \label{GenSol}
\end{equation}

This expression looks like {\it Fourier expansion} of the $u(X,t)$. So, the following Fourier expansion properties are valid for influence of the SR object: 
\[ U(\omega_n,X)=\frac{\omega_0}{2\pi}\int^{\pi/\omega_0}_{-\pi/\omega_0} u(X,t) e^{\imath \omega_n t} dt; \,\, \left\langle u^2(X,t) \right\rangle_t= \sum_n \left | U(\omega_n,X) \right |^2. \]
We can consider the expression (\ref{GenSol}) in an inertial to $K$ frame of references $K^L$, using the Lorentz coordinate transformation (without loosing the generality $X^L=(x,y,z^L)$, $\gamma = 1/\sqrt{1-\beta^2}$, $\beta=V/c$), that leads to: 
\begin{equation}
u^L(X^L,t^L)=\sum_n {\cal B}(\omega_n,X^L)e^{\imath \left[ \omega_n (\gamma-1)t^L-\omega_n \beta z^L \right]}, \label{deBroglie}
\end{equation} 
where ${\cal B}=U(\omega_n,X^L)e^{\imath \omega_n t^L}$ and (\ref{deBroglie}) looks like the {\it de Broglie} wave, corresponding to particle, if the part $e^{\imath \omega_n t^L}$ is neglected. From our point of view it means that the ``internal'' properties are neglected and are replacing with some ``macroparameters'' (energy, momentum, rest mass etc.). Together with the observed Fourier expansion properties, the Expr.(\ref{deBroglie}) may have the probabilistic interpretation. 

\section{\bf Physical Laws}
\label{sec:Law}

In previous Sections we have found some basic properties of the SR objects and SR frames. It looks like the basic properties of the SR objects correspond to the quantum mechanics ones, but the SR frames are immeasurable principally. Here we will try to find the possibility to describe the SR object and its behavior in the observer frame. 

\subsection{Omega-invariance}
Let's consider the normalized MSR with the periodical rotation matrix $A^{MSR}=A(t)= A(t+2\pi n/\omega)$, where $n$ is integer, and fix two events $(t_{1}, X'_1)$ and $(t_{2}, X'_2)$ in $K'$. One can get by the SR definition:
\begin{equation}
\Delta X'_{|2\pi n/\omega}=X'_{2}-X'_{1}=\, X_{2}A(t_{2})-X_{1}A(t_{1})=\, \Delta X \cdot A_{|2\pi n/\omega}. \label{n5}
\end{equation}
If the SR matrix is normalized ($\det A=1$), the distance between two space points in $K$ and $K'$ is found to be equal to each other:
\begin{equation}
\| \Delta X' \|^{2}_{|2\pi n/\omega}=\, \| \Delta X \|^{2}. \label{n7}
\end{equation}
For any fixed frequency $\omega$, the time points $t_{2}=t_{1}+2\pi n/\omega$ create the numerable infinite aggregate $\Lambda : \{t_{0}+2\pi n/\omega\}^{\infty}_{n=0}$ on the $t$-axis. The rotation frequency $\omega$ is the initial parameter of this aggregation. On $\Lambda$ the equality (\ref{n7}) is true, and, consequently, the interval (\ref{Interval}), as it was defined in Minkowski space, is invariant in $K$ and $K'$. We will call by definition that rotating frames $K$ and $K'$ are {\it Omega--invariant} (or $\omega${\it--invariant}). It means that SR frames on this infinite aggregate look like Lorentz-invariant and so, we will consider, that they are {\it measurable} on $\Lambda$. The frequency $\omega$ defines the {\it scale} between two frames of references $K$ and $K'$. On this approach, this parameter seems and needs to be very important in the physical object description in these frames. \\ 
Further, let's a physical object (some of its property) is described in $K'$ by the function $\psi'(X',\tau)$ and in $K$ by function $\psi(X,\tau)$. On $\Lambda$ these functions are measurable, because the interval (\ref{Interval}), as it was defined in Minkowski space, is invariant in $K$ and $K'$,  and one can write the condition expression, connecting these functions: 
\begin{equation}
\psi(X,\tau)_{|2\pi n/\omega}=\sum_l \left[ \psi'_l(X',\tau)\prod_{m_l} \exp(\pm \imath m_l \Omega\tau) \right], 
\label{SRpsi}
\end{equation} 
because for any natural $l$ and interger $m_l$, $\exp(\pm \imath m_l \Omega\tau)_{|2\pi n/\omega}=1$, where $\psi'(X',\tau)=\sum_l  \psi'_l(X',\tau)$, and in these points $X'_{|2\pi n/\omega}=X$. \\
Furthermore, we will replace $\Lambda$ by the real $t$-axis. It means that we are also replacing the discrete set on $X$ and $X'$, so that $X=X'$. Finally, we have got the expression, connecting functions $\psi$ in $K$ and $\psi'$ in $K'$ on $\Lambda$: 
\begin{equation}
\psi(X,\tau)=\sum_l \left [ \psi'_l(X,\tau)\prod_{m_l} \exp(\pm \imath m_l \Omega\tau) \right ]. \label{SRpsicom}
\end{equation} 
At any, even very high values of $\omega$, this approximation may be quite accurate, but always not complete. \\ 
Following these assumptions, we can conclude that the introduced extended ${\cal \widehat O}(3)$ group due to $\omega$--invariance hypothesis transforms to the usual ${\cal O}(3)$ group and also the Minkowski space metrics is valid for this object. This is already enough to get the basic quantum mechanics equations such as Klein-Gordon-Fock, Schr\"odinger and Dirac in the fashion standard in quantum theory \cite{Ryd85}. \\
The SR theory can explain a lot of difficulties and postulates of the quantum theory. We will illustrate it by obtaining the Klein-Gordon-Fock equation for a scalar particle without spin in the ``SR theory fashion style''. 

\subsection{KGF equation} 
For simplicity, we will analyze the case $l, m_l=1$ of Expr.(\ref{SRpsicom}):
\begin{equation}
\psi(X,\tau)=\psi'(X,\tau)\exp(\pm \imath\Omega\tau). \label{psi}
\end{equation} 
Here we have got the expression (\ref{psi}), we have already used in Sect.\ref{sec:Object} for source function (Eq.(\ref{DE})). So, it has a meaning of the SR function transformation. \\
Remind, that on $\Lambda$: $X=X'$ and we can consider $\nabla_X = \nabla_{X'}$. So, the Lorentz invariant second order differential operator, called the d'Alembertian operator, will be invariant in frames $K$ and $K'$: 
\[ \Box =  \frac{\partial^2 }{\partial \tau^2}-\nabla^2_X = \frac{\partial^2 }{\partial \tau^2}-\nabla^2_{X'} = \Box'. \]
We are able to apply the d'Alembertian operator to both parts of the Eq.(\ref{psi}). It will lead to equation: 
\begin{eqnarray}
\frac{\Box \psi(X,\tau) }{\psi(X,\tau)} =\frac{\Box' \psi'(X',\tau) }{\psi'(X',\tau)} -\Omega^2 + \frac{2 \imath \Omega }{\psi'(X',\tau)}\frac{\partial \psi'(X',\tau)}{\partial \tau}.  \label{KGF0} 
\end{eqnarray} 
The third term right may be neglected in comparison with the second one in case of the stable particle, stable, at least, in comparison with the period of rotation $T=2\pi /\omega$: $\Omega \gg \left |(\partial \psi' /\partial \tau)/\psi' \right |$. If the physical object in $K'$ (it may be, for example, an electromagnetic wave) is satisfied the equation $\Box' \psi'(X,\tau)=0$, then supposing $\Omega = mc/\hbar$, one can get from (\ref{KGF0}) the {\it Klein-Gordon-Fock equation}: 
\begin{equation}
 \frac{\Box \psi }{\psi}+ \frac{m^2 c^2}{\hbar^2} =0 . \label{KGF}
\end{equation}
The equation (\ref{KGF}) is valid for any inertial to $K$ frames of references due to the Lorentz invariance of the d'Alembertian operator, that was directly shown in \cite{NB01}. There was also shown the obtaining of the Schr\"odinger equation in these papers, which is the non-relativistic approximation to the Klein-Gordon equation. As far as the $\omega$--invariance gives the necessary symmetries, it is possible to obtain the Dirac equation by the usual way \cite{Ryd85}. Note, that (\ref{n7}) is not satisfied for SSR, so equations, created for SSR, apparently, will correspond to the QCD, QFT equations. \\
From the SR point of view, the equation (\ref{KGF}) contains a new idea -- an idea of connecting the physical object properties in different rotating frames of references. It is a consequence of the declared SR equivalence principle. \\
We may conclude, that on numerable infinite aggregate $\Lambda$ (discrete set), SR objects are both Lorentz- and Omega- invariant. So they looks like the quantum physics objects and also satisfy to special relativity. This principally coincide with the conclusion of V.Petkov \cite{Petk02}, about the work of A.Anastasov \cite{Anast93} of the possible ``structure'' of the quantum mechanics objects to correspond to the special relativity. Now we can see why. \\

$\bullet${\it The ``Einstein's formula'' for the SR} \\
The invariant $E^2/c^2-\vec{p}\cdot\vec{p}=m^2c^2$ given by the energy-momentum 4-vector of a particle $p^{\mu}=\left({E}/{c}, \vec{p}\right)$ in quantum mechanics corresponds to the KGF equation. An electromagnetic wave in $K'$ gives for this invariant $E'^2/c^2-\vec{p'}\cdot\vec{p'}=0$. For the stable and localized object in $K$ ($\vec{p}=\vec{0}$) this invariant will be $E^2/c^2=m_0^2 c^2$, so from (\ref{KGF0}) we can get the well-known Einstein's formula: 
\begin{equation}
E^2 = C(\Omega)c^2 = m_0^2(\Omega)c^4. \label{E-mc}
\end{equation}
Note that the ``influence'' of the SR object (Sect.\ref{sec:Object}) may be interpreted as an additional mass to (\ref{E-mc}), but it has another origin. Here we can see some correspondence of the rotation frequency and the rest mass. This question needs to be analyzed in details, but not in this paper (see also conclusions in Sect.\ref{sec:Conclusion}). \\ 

\section{\bf Space-Time Properties}
\label{sec:Properties}

The equivalence principle of the Space rotation theory leads to some new and interesting aspects of our understanding of Space-time and its properties.\\
\begin{itemize}
\item Mathematically, any SR, considered in this paper, at any fixed time $t$ (time is only a parameter of this transformation), transfers the ${\cal R}^3$ space into itself. In special relativity time and space form a specific space-time as a whole, with own metrics and properties. That's why the uniform transformation in mathematics leads to some paradoxes in physics. We have seen above, that the ``physical space'' of $K'$ consists of the internal, visible space and external, invisible one from the $K$ point of view. So, visible space of $K'$ is included to the visible space of $K$. Because of the SR equivalence, considering $K$ from the $K'$ point of view, one may conclude, that the visible space of $K$ is included to the visible space of $K'$. This is a physical paradox that may be considered as a new type of the space-time topology. 
\item The considered SR intervals corresponds to the metric tensor with non-diagonal elements (even averaged in time), so, generally, the rotating frames are immeasurable. The introduced Omega--invariance hypothesis lets make them measurable, but, of course, with some limitations. These limitations are in a good agreement with the quantum physics postulates and paradoxes, that gives us some confidence on our way. We keep in mind that these limitations are coming from the SR theory; they are not postulated or introduced. We may suppose, that ``new'' interactions, declared in quantum physics, such as strong or electroweak, are reflections of the ``usual'' forces, existing in rotating frames, to the observer frame. This is a way to the Unification theory of these forces. 
\item The correspondence of the SR object properties and quantum physics ones gives us another confidence of the legitimacy of the SR theory. For example, the quantification, spin etc. are essential parts of this theory, we do not need to declare something new, to introduce these object properties. As far as SR object is confined inside a very specific region -- the horizon surface, some SR object characteristics, like spin for example, look like real ``internal'', ``hidden'' parameters. It corresponds to real situation. Also, the SR object types, we can imagine, are in a good agreement with known quantum physics objects. The quantification conditions correspond to the energy conservation law in the observer's frame. It is also object stability conditions. This way, such unstable elementary particles as resonancies may be interpreted as the SR objects, for which the quantification conditions could not be satisfied in time period, defined by $\omega$--invariance.  
\end{itemize}
At last, we would like to consider some philosophical consequences of the Space rotation theory and its basic principle -- the space rotation equivalence. After declaring this principle, one can ask a lot of questions, such as: are where an infinite set of ``parallel'' ``rotating worlds'', and, if there are, so why do not we observe them in the everyday life and, is it possible, for example, to get to ``another world'' by rotating in centrifuge etc. \\

$\bullet$ {\it Rotating World.} \\
The observer is connected with some PhRD, so he is connected with all included components. If we consider a real man, as an observer, he is a part of the PhRD and consists of the physical objects, existing in this PhRD. On the other hand, the observer creates the PhRD, it is his product, as a child builds his own world after his birth. He builds his own ``projection'' of the real World. In another SR frame, it will be another physical objects, so the observer in one SR frame does not exist in other one as a physical object. But it does not mean that there are no any other ``observers'' in another SR frames. So, the question of existence of the ``parallel rotating worlds'' is open. It may be that our visible Universe is only one projection of the real Universe to our physical reality description system (Platon has said something similar long time ago). \\

$\bullet$ {\it Physical vacuum.} \\
Another possibility is that our Universe is a result of a spontaneous symmetry breakdown (in ``macro-scale'') of the real Universe, so this symmetry is observable only in ``micro-scale'', the scale of the elementary particles, vacuum etc. The difference between particles and vacuum is that physical vacuum is unrealized particles in observer's SR frame. In other words, vacuum looks like unstable particles that cannot be identified as stable SR objects. Such representation also coincides with modern understanding of the physical vacuum as a lot of virtual particles. \\

We can observe the ``consistence'' of other rotating frames only in frames of $\omega$--invariance, so all our experiments will give principally incomplete information. 

\section*{\bf Conclusions and Remarks}
\label{sec:Conclusion}

It seems that our attempt to create the system of the physical reality description applying to space rotations, undertaken in this paper, was successful enough and we may declare the result as a Space rotation theory. The SR theory is based on the SR equivalence, the quantification principle and the $\omega$--invariance hypothesis. \\
The SR equivalence principle is a generalization of the Relativity principle by the extension it to the space rotation frames. It includes the special relativity principle as a particular case of the Equivalence principle. Also, the Equivalence principle extends the conception of the space symmetry of the ${\cal O}(3)$ space rotation groups. This extended rotation group ${\cal\widehat O}(3)$ adds some very important physical characteristics to space-time and SR object descriptions. The stable-state objects, correlated with known quantum physics objects, are found. \\
The quantification principle is based on the usual energy conservation law for found SR objects. It does not need any additional postulates to explain the quantum properties of existing particles and fields. \\
The $\omega$--invariance hypothesis makes SR frames measurable, but, of course, with some limitations. These limitations are coming from the SR theory; they are not postulated or introduced. They are in a good agreement with the quantum physics postulates and paradoxes. \\
From the SRT point of view, quantum physics is an effort of the approximate description of the immeasurable SR systems, because in quantum physics the numerable infinite aggregate on $t$-axis is replaced by the continuous $t$-axis. At any, even very high values of $\omega$, this approximation may be quite accurate, but always not complete. It looks like the reason of the {\it uncertainties} in quantum physics, its incompleteness and formalism. We may suppose, that ``new'' interactions, declared in quantum physics, such as strong or electroweak, are reflections of the ``usual'' forces, existing in rotating frames, to the observer's frame and also ``influence'' of the SR objects in our frame. This is a way to the Unification of these forces. \\
All these facts make us declare the ``initial'', basic origin of the SR theory in comparison with the quantum physics. Also, on other hand, the invariance of the special relativity looks like a particular case of the space rotation Equivalence, the basic principle of SRT. So, SRT includes both special relativity and quantum physics basic principles, really let them fit together! \\ 
We have not considered the general relativity in this paper, because it is not completely clear to us, what does gravitation mean from the SRT point of view. But we believe that some correspondences mentioned above of rotation frequencies, potentials with the rest mass give us a way to further investigations of this question. 
%
%


\begin{thebibliography}{99}
\bibitem{GM81} M.Gell-Mann. Questions for the Future./in: The Nature of Matter, Wolfson College Lectures 1980.-- Wolfson College, Oxford 1981.
\bibitem{Butt99} J.Butterfield. The State of Physics: 'Halfway through the Woods'. -- Journal of Soft Comp., IQS Associat., 1999. 
\bibitem{AE05} A.Einstein. The Principle of Relativity. -- Methuen and Company, Ltd. of London, 1923 (1905).
\bibitem{AE35} A.Einstein, B.Podolsky, N.Rosen. Can Quantum-Mechanical Description of Physical Reality be considered Complete? -- Physical Review 47, pp. 777, 1935. 
\bibitem{Bell64} J.S.Bell. On the Einstein Podolsky Rosen Paradox. -- Physics 1, No.3, pp.195, 1964. 
\bibitem{Asp82} A.Aspect et al. Experimental realization of Einstein-Podolsky-Rosen-Bohm gedankenexperiment: A new violation of Bell's inequalities. -- Physical Review Letters 49, No.2, p.91, 1982 // Experimental test of Bell's inequalities using time-varying analyzers. -- Physical Review Letters 49, No.25, p.1804, 1982. 
\bibitem{Chi93} R.Y.Chiao, P.G.Kwiat, A.M.Steinberg. Faster than light?. -- Scientific American, Aug. 1993.
\bibitem{Ryd85} L.H.Ryder. Quantum Field Theory.-- Cambridge Univ. Press, 1985. 
\bibitem{Ryd01} L.H.Ryder, B.Mashhoon. Spin and Rotation in General Relativity. -- {\bf gr-qc/0102101}, February, 2001. 
\bibitem{Mash00} B.Mashhoon. Reality and Nonlocality.-- {\bf gr-qc/0011013}, November, 2000. 
\bibitem{NB03} A.V.Novikov-Borodin. Space Rotation.-- The V-th Int.Conf. ``Symmetry in Nonlinear Mathematical Physics'', Kiev, Ukraine, June 23-29, 2003./ Proceedings of Institute of Mathematics of NAS of Ukraine, 2004 (to be published). 
\bibitem{NB01} A.V.Novikov-Borodin. Space Rotation Invariance.-- DESY Report M 01-02, May, 2001// {\bf quant-ph/0105011}, May, 2001. 
\bibitem{Log89} A.A.Logunov, M.A.Mestvirishvilli. Relativistic Theory of Gravitation.-- Nauka, Moscow 1989. 
\bibitem{Poin04} H.Poincar\'e //Bull.de Sciences Math.-- vol.28, ser.2, p.302, 1904; The Monist.-- vol. XV, p.1, 1905. 
\bibitem{Land85} L.D.Landau, E.M.Lifschitz. Theoretical Physics: The Classical Theory of Fields, vol.2.-- Nauka, Moscow, 1988. 
\bibitem{Liu00} Y.Liu. Geometric origin of the weak CP phase.-- Phys. Review D, Vol.61, 033010, 2000.
\bibitem{Vlad81} V.S.Vladimirov. Mathematical Physics Equations.-- Nauka, Moscow 1981.
\bibitem{Petk02} V.Petkov. Acceleration-dependent electromagnetic self-interaction effects as a basis for inertia and gravitation.-- {\bf physics/9909019}, Sept. 2002. 
\bibitem{Anast93} A.H.Anastasov. Annuaire de l'Universite de Sofia St. Kliment Ohridski, Faculte de Physique 81, 135, 1993.// A.H.Anastassov. The Theory of Relativity and the Quantum of Action (4-Atomism). -- Doctoral Thesis, Sofia University, Sofia, 1984 (unpublished).


\end{thebibliography}
\end{document}